\documentclass[12pt]{article}
\usepackage{amsfonts,amssymb}

\newcommand{\la}{\lambda}

\newcommand{\pa}{\partial}

\newcommand{\be}{\begin{equation}}
\newcommand{\ee}{\end{equation}}
\newcommand{\beq}{\begin{eqnarray}}
\newcommand{\eeq}{\end{eqnarray}}
\newcommand{\ma}{$r$-matrix }

\newcommand{\beqq}{\begin{eqnarray*}}
\newcommand{\eeqq}{\end{eqnarray*}}

\newcommand{\R}{\mathbb{R}}

\newcommand{\C}{\mathbb{C}}

\newcommand{\Z}{\mathbb{Z}}

\title{\Large \bf Category of nonlinear evolution equations,
algebraic structure, and \ma}
\author{Zhijun Qiao\\
{\small Theoretical Division and Center for Nonlinear Sciences}\\
{\small Los Alamos National Laboratory, Los Alamos, NM 87545, USA}\\
{\small Institute of Mathematics, Fudan University, Shanghai 200433, PR China}\\
{\footnotesize  E-mails: qiao@cnls.lanl.gov \ \ \ qiaozj@hotmail.com}\\
     Cewen Cao\\
{\small Department of Mathematics, Zhengzhou University, Zhengzhou 450052,
       P. R. China}\\
{\footnotesize  E-mail: cwcao@public2.zz.ha.cn}\\
      Walter Strampp\\
{\small Fachbereich 17 - Mathematik/Informatik, Universit\"at-GH Kassel}\\
{\small Heinrich-Plett-Str. 40, D-34109 Kassel, Germany}\\
{\footnotesize  E-mail: strampp@hrz.uni-kassel.de}}
\date{}
\begin{document}
\maketitle
\begin{abstract}
This paper deals with the 
category of nonlinear evolution equations (NLEEs) 
associated with the spectral problem
and provides an approach for constructing their
algebraic structure and $r$-matrix.
First we introduce the category of  NLEEs, which is composed of 
various positive order and negative order
hierarchies of NLEEs both integrable and non-integrable.
The whole category of NLEEs possesses a generalized Lax representation.
Next, we present two different Lie algebraic structures
of the Lax operator, one of them is universal in the category,
i.e. independent of the hierarchy, while the other one is
nonuniversal in the hierarchy, i.e. dependent on the underlying hierarchy.
Moreover, we find that
two kinds of adjoint maps are $r$-matrices under
the algebraic structures. In particular, the Virasoro algebraic
structures without central extension
of isospectral and non-isospectral Lax operators
can be viewed as reductions of our algebraic structure. Finally,
we give several concrete examples to illustrate our methods.
Particularly, the Burgers category is linearized
when the generator, which generates the category, is chosen to be
independent of the potential function. Furthermore, an isospectral
negative order hierarchy in the Burger's 
category is solved with its general solution.
Additionally, in the KdV category we find an interesting fact:
the Harry-Dym hierarchy is contained in this category as well as 
the well-known Harry-Dym equation is included in a positive order
KdV hierarchy.
\end{abstract}


$ $\\
{\bf AMS Subject: 35Q53, 58F07, 35Q35.}
\section{\bf Introduction}

Integrability study of nonlinear evolution equations
has been an attractive topic in soliton theory and nonlinear
phenomenon. Calogero \cite{C1} proposed the $C$-integrable
(namely, linearizable by an approriate Change of variables) 
and $S$-integrable (namely, integrable via some spectral transform
technique) terminology for dealing with nonlinear partial diiferetial
equations (PDEs). Many nonlinear PDEs were shown $C$-integrable and
$S$-integrable \cite{CJ1}. Mikhailov, Shabat and  Sokolov \cite{MSS1}
discussed some classes of nonlinear $C$-integrable and
$S$-integrable PDEs through using the symmetry
approach.
Flaschka, Newell and Tabor \cite{FNT1} considered in detail
the Painleve analysis process  for both ODEs and PDEs
and investigated its test for integrable equations.
 
On the other hand, the $r$-matrix method 
is also an important part in classical and quantum integrable
systems \cite{1}. The classical $r$-matrix has been first introduced 
by Sklyanin in Refs.  \cite{2} and  \cite{3} as the limit of its quantum counterpart.
Subsequently, Drinfeld used this to introduce a new geometric notion, that of
a Poisson Lie group \cite{4}. Following Drinfeld's ideas \cite{4} Semenov-Tian-Shansky
showed that the concept of classical $r$-matrix leads to an algebraic
construction of integrable systems generalizing the AKS scheme.
In terms of the $r$-matrix \cite{5} an effective view
of the multi-Hamiltonian property of such equations can be presented.
In addition, it gives a general explanation of the dressing transformations
used for obtaining solutions in terms of group factorizations \cite{6}.
In Ref. \cite{7} Jimbo constructed explicit solutions of the quantum
YB equation for the generalized Toda system
and moreover obtained many beautiful results \cite{8,9,10}
by using the $r$-matrix method.

For the study of algebraic structure of
integrable evolution equations, there has been also
discussion in the literature. For example, the well-known
$W$-algebra was constructed by Orlov and Schulman
through using the vertex operator \cite{OS1}.
The KP system was also found to have this kind of $W$-algebraic
structure by Dickey \cite{Dickey1}, which includes the Virasoro algebra
as its subalgebra. The $W$-algebra played an important role
in the the so-called second Poinsson structure \cite{Dickey1}. For this,
the most important is to find the generators of $W$-algebra.
All these facts were only for the case of integrable hierarchies.
How about the case for both integrable and non-integrable hierarchies?
This paper will deal with this problem through introducing
the category of nonlinear evolution equations (NLEEs).
The category of NLEEs  develops the positive order  to the negative 
order hierarchies for both  the integrable 
and the non-integrable cases. In particular,
the positive and the negative order integrable hierarchies 
 will be generated by the recursion operator, its inverse, and some kernel
 elements from the pair of Lenard's operators. 
Mikhailov,  Shabat and  Sokolov \cite{MSS1} ever extended the integrable
equations
 by employing the symmetry procedure and discussed the classifications
for the integrable hierarchies. All of their results were for
$C$-integrable and $S$-integrable case. In this paper, we will discuss the case
for both integrable and non-integrable hierarchies and will not
interfere the existence of symmetries. 
Here, we point out that
 throughout this paper: ``integrable'' 
means the sense of Lax, namely, the PDE admits isospectral
(i.e. $\la_t=0$)
or usual non-isospectral (i.e. $\la_t=a\la^n, \ n\in \Z, \ a\in \R
/\C$) Lax form; otherwise, we say the PDE is non-integrable
in the sense of Lax form.

The purpose of the present paper is to give an approach to the category of 
nonlinear evolution equations directly from a spectral problem
and to connect the $r$-matrix to the category of NLEEs. The whole
paper is organized as follows.
In the next section we first introduce the notation of category of NLEEs,
which is composed of various positive and negative order hierarchies of
both integrable and non-integrable NLEEs, and then we give
the generalized Lax representation (GLR). 
In sections 3 and 4 we respectively present two
different Lie algebraic structures of the Lax operator.
One structure is produced independent of the hierarchy in the category
while the other holds only within one hierarchy.
Moreover, by using these algebraic structures we find that
two kinds of adjoint maps result in $r$-matrices for the NLEEs.
In section 5, it is pointed out that the well known Virasoro algebraic
structures (without the central extension) of isospectral and non-isospectral Lax operators
are obtained as reductions of our algebraic structure.
Finally, in section 6 the examples of
several continuous spectral problems
are given to illustrate our methods.
Particularly, the Burgers category is linearized
when the generator, which generates the category, is chosen to be
independent of the potential function. Furthermore, an isospectral
negative order hierarchy in the Burger's category is solved with its general solution.
Additionally, in the KdV category we find an interesting fact:
the Harry-Dym hierarchy is contained in this category as well as
the well-known Harry-Dym equation is included in a positive order
KdV hierarchy.

     Before displaying our main results, let us first give some
     necessary notation:

    $x\in R^{l}$, $t\in R$, $u=(u_{1},...,u_{m})^{T}\in S^{m}(R^{l},R)=
\overbrace{S(R^{l},R) \times \cdots \times S(R^{l},R)}^{m}$, $u_{i}=u_{i}(x,t)\in S(R^{l},R)$, $i=1,2,...,m$,
for arbitrarily fixed $t$, $S(R^{l},R)$ stands for the Schwartz function space
on $R^{l}$. ${\cal B}$ denotes all complex (or real) value functions $P(x,t,u)$
of the class $C^{\infty}$ with respect to $x,t$, and of the class $C^{\infty}$
in Gateaux's sense with respect to $u$.
${\cal B}^N=\{(P_1,...,P_N)^T\vert P_i\in {\cal B}\}$, ${\cal V}^N$
stands for all linear operators $\phi=\phi(x,t,u)$: ${\cal B}^N \rightarrow {\cal B}^N$
which are of the class $C^{\infty}$ with respect to $x, t$, and
of the class $C^{\infty}$ in Gateaux's sense with respect to $u$.


    The Gateaux derivate of vector function $X\in {\cal B}^n$ in
the direction $Y\in{\cal B}^m$ is defined by
\begin{equation}
X_{*}(Y)=\frac{d}{d\epsilon}\vert_{\epsilon=0}X(u+\epsilon Y).
\end{equation}
For the two arbitrary vector fields $X, Y\in {\cal B}^m$, define
the following operation
\begin{equation}
[X, Y]=X_{*}(Y)-Y_{*}(X).
\end{equation}
Then, ${\cal B}^m$ composes a Lie algebra about the above multiplication
operation \cite{11}.   For the operator $\phi \in {\cal V}^N$, its Gateaux
derivate operator $\phi_{*}: {\cal B}^m \rightarrow {\cal V}^N$ in
the direction $\xi$ is defined as follows
\begin{equation}
\phi_{*}(\xi)=\frac{d}{d\epsilon}|_{\epsilon=0}\phi(u+\epsilon\xi), \quad \xi\in {\cal B}^m.
\end{equation}


    If not otherwise stated, the spectral operators $L=L(u)$ (or the spectral
operators $L=L(u,\lambda)$ with the spectral parameter $\lambda$) considered in
this paper are denoted by $L\in{\cal V}^N$, and we always assume that $L_{*}:
{\cal B}^m\rightarrow{\cal V}^N$ is an injective homomorphism. An operator $H$ acting on 
a function $f$ is denoted by $H\cdot f$. $I$ stands for the $N\times N$ unit 
operator.


\setcounter{equation}{0}
\section{Category of NLEEs and generalized Lax representation (GLR)}
    In this section, a procedure for constructing the category of NLEEs and generalized 
Lax representations are presented, and moreover
it is shown how to construct the $L-A-B$ triple representation \cite{12}
for a given nonlinear quation.


   Let us start from 
a general $N\times N$ spectral problem
\begin{equation}
 L\cdot\psi=\la\psi, \quad L\in {\cal V}^N,
\end{equation}
where $\lambda $ is a spectral parameter, $\psi \in {\cal B}^N$. Denote
the functional gradient of spectral parameter $\lambda$ with regard to
the potential vector $u$ by
$\frac{\delta \lambda}{\delta u}=
(\frac{\delta \lambda}{\delta u_1},\cdots,\frac{\delta \lambda}{\delta u_m})^T$.
Tu and Cao respectively gave some discussions
about the calculations of the functional gradient
in Ref. \cite{13} and Ref. \cite{14}. Strampp ever studied
recursion operators, spectral problems, and B\"acklund
transformations by introducing a relation between recursion operators
and eigenvalue functions \cite{15, 16}. Thus, we define the
Lenard operators as follows.

    {\bf Definition 2.1} \quad {\it If there exists a pair of
$m\times m$ operators $K=K(u),J=J(u):S^m(R^l,R)\rightarrow S^m(R^,R)$ such that
\begin{equation}
K\cdot\frac{\delta \lambda}{\delta u}=\la^c  J\cdot\frac{\delta \lambda}{\delta u},
\end{equation}
then $K,J$ are called a pair of Lenard operator of (2.1),
and (2.2) is called the Lenard spectral problem of (2.1). Here the constant
$c$ is definitely chosen by the concrete form of (2.1)}.


    In many cases, there exist (but not unique) the pair of Lenard's operators
satisfying (2.2), and frequently both of them are Hamiltonian operators. For
instance, for the KdV-Schr$\ddot{\rm o}$dinger spectral problem
 $\psi_{xx}+u\psi=\lambda \psi$, $\frac{\delta \lambda}{\delta u}=\psi^2$, only
choosing $K=-\frac{1}{4}\partial^3
-\frac{1}{2}(u\partial+\partial u)$, $J=\pa=\frac{\pa}{\pa x}$, we have $K\cdot\frac{\delta \lambda}{\delta u}
=\la  J\cdot\frac{\delta \lambda}{\delta u}$. Eq. (2.2) plays an
important role in the nonlinearization theory and the construction of completely
integrable finite-dimensional systems \cite{17}.


    Let $M=(m_{ij})_{N\times N},\tilde{M}=(\tilde{m}_{ij})_{N\times N}$
be the arbitrarily given $1+l$ dimensional (i. e. independent
variables $(x,t)\in R^l\times R, \ l\geq 1$) linear $N\times N$
matrix operators.
Then we have the following definitions.


   {\bf Definition 2.2} \quad {\it $G_0\in S^m(R^l, R),G_{-1}\in S^m(R^l, R)$
are respectively called the positive order and the negative order 
generators, if they respectively satisfy the operator equations}
\begin{eqnarray}
  L_{*}(J\cdot G_0) &=& M,\\
L_{*}(K\cdot G_{-1})&=& \tilde{M}.
\end{eqnarray}


    Denote the solution sets of (2.3) and (2.4) by ${\cal N}_J(M)$ and
${\cal N}_K(\tilde{M})$, respectively. In general, they are not empty.


    {\bf Definition 2.3} \quad {\it  Let ${\cal N}_J(M)\ne\emptyset,
{\cal N}_K(\tilde{M})\ne\emptyset$ and choose $G_0\in {\cal N}_J(M),
G_{-1}\in {\cal N}_K(\tilde{M})$. Write the recursion operator
${\cal L}=J^{-1}K$. The sequence
$\{G_j\}^{\infty}_{j=-\infty} \subseteq S^m(R^l,R)$
recursively determined by
\begin{equation}
G_j=\left\{\begin{array}{ll}
{\cal L}^j\cdot G_0, & j\geq 0,\\
{\cal L}^{j+1}\cdot G_{-1}, & j<0,
\end{array}
\right.
\end{equation}
is called the Lenard's sequence of (2.1);
the set of the following nonlinear equations
\begin{equation}
u_t=X_m(u,G_0,G_{-1}), \ m\in Z,
\end{equation}
produced by the vector field
\begin{equation}
X_m(u,G_0,G_{-1})\stackrel{\triangle}{=}
J\cdot G_m, \ m\in Z,
\end{equation}
is called the category of nonlinear evolution equations of (2.1).
The subset of the equations (2.6) obtained for $m\ge 0$
is called the positive order category while the subset
obtained for $m<0$ is called the negative order category.}

    Apparently, the positive and the negative order generators
$G_0,G_{-1}$ depend on the choice of matrix operators $M,\tilde{M}$,
thus the category (2.6) is composed of various hierarchies (both integrable
and non-integrable) of NLEEs
which are generated according to the choice
of operators $M,\tilde{M}$.

For example,
with $M\equiv 0$ (i.e. $G_0\in Ker \ J$), the hierarchy in the positive order category of (2.6) just reads as the isospectral hierarchy of evolution equations
\cite{18}; with $\tilde{M}\equiv 0$ (i.e. $G_{-1}\in KerK$), the hierarchy in the 
negative order category of (2.6) is exactly the second isospectral hierarchy of
evolution equations studied in Ref. \cite{19}.
Additionally, the negative order generator $G_{-1}$ can be considered
to produce finite-dimensional constrained
Hamiltonian systems \cite{20}. Obviously, the negative order category of
(2.6) is generated with the help of the inverse recursion operator
${\cal L}$. Strampp and Oevel gave the inverse recursion operator
in an explicit form for the nonlinear derivative Schr\"{o}dinger
equation \cite{21}. In 1991 we suggested the commutator
representations for the negative order hierarchy
of isospectral NLEEs \cite{23}. Afterwards, we \cite{24}
further found that the
same spectral problem can generate two different hierarchies of
integrable NLEEs, one is the usual higher order (i.e. positive order)
hierarchy of NLEEs, the other is the negative order hierarchy of NLEEs.
All these equations have the Lax representations \cite{24}.
Here
we study the generalized case, i.e. the category of NLEEs.

With $M=I$ or $\tilde{M}=I$, under the basic
condition ${\cal N}_J(I)\ne\emptyset$ or
${\cal N}_K(I)\ne\emptyset$ Eq.
(2.6) actually gives the positive and the negative order hierarchies
of non-isospectral evolution equations, which can be obtained from the
following Theorem 2.2.
Thus, by the arbitrariness of $M$ and $ \tilde{M} $, Eq. (2.6) unifies
together all possible hierarchies of evolution equations associated
with the spectral problem (2.1).
Due to this fact, Eq. (2.6) is named
`the category of nonlinear evolution equations'.


    {\bf Theorem 2.1}\quad {\it Let $M=(m_{ij})_{N\times N},
    \tilde{M}=(\tilde{m}_{ij})_{N\times N}$
be two arbitrarily given $N\times N$ linear matrix operators,
${\cal N}_J(M)\neq \emptyset$, and ${\cal N}_K(\tilde{M})\neq \emptyset$.
Suppose that for $G=(G^{[1]},\cdots,G^{[m]})^T\in S^m(R^l,R)$
and $\alpha,\beta\in Z$ the operator equation
\begin{equation}
  [V, L]=L_{*}(K\cdot G)L^{\beta} - L_{*}(J\cdot G)L^{\alpha}
\end{equation}
possesses a solution $V=V(G)$,
then the vector field $X_m=X_m(u,G_0,G_{-1})$
satisfy
  \begin{equation}
   L_{*}(X_{m})=\left[W_m,L\right]+\bar{M}L^{m\eta}, \ m\in Z, \ \
   \bar{M}=\left\{\begin{array}{ll}
   M, & m\geq0, \\
   \tilde{M},& m<0,
   \end{array}\right.
\end{equation}
where $\eta=\alpha-\beta$ and the operator $W_m$ is given by
  \begin{equation}
    W_{m}=\sum V(G_{j})L^{(m-j)\eta-\alpha}, \ \
    \sum=\left\{\begin{array}{ll}
   \sum_{j=0}^{m-1}, & m>0, \\
   0, & m=0,\\
   -\sum_{j=m}^{-1}, & m<0. \end{array}\right.
\end{equation}
Here $G_j$ are determined by (2.5), and $L^{-1}$ is the inverse of $L$,
i.e.  $LL^{-1}=L^{-1}L=I$, and $[\cdot,\cdot]$ denotes the usual
commutator.}

          {\bf Proof}\quad For $m=0$, it is obvious.
          For $m>0$,
      \begin{eqnarray*} \left[W_m,L\right]&=&
                 \sum^{m-1}_{j=0}\left[V(G_j),L\right]L^{(m-j)\eta-\alpha}\\
        &=&\sum^{m-1}_{j=0}\{L_{*}(K\cdot G_j)L^{(m-j-1)\eta}-L_{*}(J\cdot G_j)L^{(m-j)\eta}\}\\
        &=&\sum^{m-1}_{j=0}\{L_{*}(J\cdot G_{j+1})L^{(m-j-1)\eta}-L_{*}(J\cdot G_j)L^{(m-j)\eta}\}\\
        &=&L_{*}(X_m)-L_{*}(J\cdot G_0)L^{m\eta}\\
             &=&L_{*}(X_m)-ML^{m\eta}. \end{eqnarray*}
    For $m<0$, the proof is similar. \hfill \rule{2mm}{4mm}


    {\bf Remark 2.1} \quad  The structure equation (2.8) of
commutator representations is a natural generalization of the structure
equation
$[V,L]=L_{*}(K\cdot G)-L_{*}(J\cdot G)L$ presented by Cao Cewen \cite{25}.


    {\bf Remark 2.2} \quad  The choice of constants $\alpha, \beta \in Z$
is determined by the concrete form of (2.1).
    In many cases \cite{24}, $V=V(G)$ can be
solved for the given $L$.


    {\bf Theorem 2.2} \quad { \it The category (2.6) of NLEEs
has the following representation}
\begin{eqnarray}
L_t=\left[W_m,L\right]+\bar{M}L^{m\eta}, \ m\in Z, \ \
   \bar{M}=\left\{\begin{array}{ll}
   M, & m>0, \\
   \tilde{M},& m<0.
   \end{array}\right.
\end{eqnarray}


    {\bf Proof} \quad For $m\geq 0$, because $L_{*}(u_{t})=L_{t}$
    and $L_{*}$ is injective,
$$  L_{t}=\left[W_m,L\right]+ML^{m\eta} \Longleftrightarrow L_{*}(u_{t}-X_m)=0
    \Longleftrightarrow u_{t}=X_m.  $$
which completes the proof. \hfill \rule{2mm}{4mm}


    {\bf Definition 2.4}\quad { \it  Eq. (2.11) and $W_m$ are called
the generalized Lax representations (GLR) and the generalized Lax-operator
(GLO), respectively.}


Obviously, with $\bar{M}=0$ (i.e. $G_0\in Ker J,\ G_{-1}\in Ker K$), Eq. (2.11) reduces
the standard (i.e. isospectral case: $\la_t=0$)
Lax representations, and with $\bar{M}=I$ (of course ${\cal N}_J(I)\ne\emptyset$ and
${\cal N}_K(I)\ne\emptyset$ are needed), Eq.
 (2.11) reduces the non-isospectral
(i.e. $\la_t=\la^{m\eta},\ m\in Z$)
Lax representations. For two special cases:
the isospectral case (i.e. $M=\tilde{M}=0$) and the non-isospectral case
(i.e. $M=\tilde{M}=I$),
Ma \cite{26} discussed the Lax operator algebras of the positive order
(i.e. $m>0$) hierarchy of NLEEs. But a general framework
has not been obtained for
all integer $m\in Z$ and all linear
matrix operators $M, \tilde{M}$.
In the following sections, we shall construct a general frame --
generalized algebraic structure and furthermore present the 
$r$-matrix
for the category of NLEEs.

    {\bf Remark 2.3} \quad Eq. (2.11) admits the
structure of $L-A-B$ representations of the category (2.6)
in an explicit form.
Thus, we give a constructive approach to the Manakov operator pair
{\it A, B} in the $L-A-B$ triple representation \cite{12}. In
Ref. \cite{QS},
we determined the range of the $L-A-B$ triple representation through
defining the Lie quotient algebras.

    {\bf Remark 2.4} \quad Eq. (2.11) contains both the integrable
    and the non-integrable hierachies because of the multiple
    choices of $\bar{M}$. Therefore, our category of NLEEs
    are not included in the system of multi-component
    KP and its reduction.

    {\bf Corollary 2.1} \quad
Assume that the potential vector function $u$ is independent of $t$
and the following condition holds
$$
 [\sum_{i=-r}^sc_iW_i, L]=-\bar{M}\sum_{i=-r}^{s}c_iL^{i\eta}
$$
with constants $c_i$, $(-r\leq i\leq s)$.
Then $u$ will satisfy the stationary system of the category (2.6)
$$
\sum_{i=-r}^sc_iX_i(u)=0,\  \forall r,s\in Z^{+}.
$$

We shall give several concrete examples in section 6.

\section{Universal algebraic structure and \ma}
\setcounter{equation}{0}
From (2.9), we have seen that for various linear
matrix operators $M, \tilde{M}$,
the category (2.6) of NLEEs indeed yields 
different hierarchies of NLEEs.
That means, the hierarchy in the category (2.6) 
changes according to the choice of $M, \tilde{M}$.
In this section, we shall construct the algebraic structure
and $r$-matrix which holds for all hierarchies of NLEEs
in the category (2.6).
Let us start from the following definition.


    {\bf Definition 3.1} \quad
{ \it
Suppose that for a spectral operator $L\in {\cal V}^N$
and an integer $n\in Z$ there exist pairs $(A,M)$ of
vector fields $X\in {\cal B}^m$
and operators $A, M\in {\cal V}^N$ with the property
\begin{equation}
[A,L]=L_{*}(X)-ML^n.
\end{equation}
Then $(A,M)$ is called a Manakov operator pair of $L$.
The set of all Manakov operator pairs is denoted by
${\cal M}^n_L$. $X$ is called the vector field
corresponding to $(A,M)$. The set of all vector fields $X$
is denoted by $V({\cal M}^n_L$). The set of all triples
$(A,M,X)$ is denoted by ${\cal P}^n_L$.}


As long as Eq. (2.8) has an operator solution
for a given $L\in {\cal V}^N$, then by theorem 2.1 and
Eq. (2.9) there exists a triple
$(A,M,X)\in{\cal P}^n_L$ satisfying (3.1).


It is easy to prove the following proposition.


    {\bf Proposition 3.1}


    { (i) \it \quad
The vector field associated with each
Manakov operator pair is unique;


\rm (ii) \it \quad both ${\cal P}_L^n$ and ${\cal M}_L^n$ 
form linear spaces.}


Apparently, if there is $A,M\in {\cal V}^N$ for $X\in {\cal B}^m$
such that Eq. (3.1) holds, then $u_t=X$ possesses the GLR
$L_t=[A,L]+ML^n$.
It is not difficult to see that ${\cal P}^n_L$ and ${\cal P}^0_L$, ${\cal M}^n_L$ and ${\cal M}^0_L$
are equivalent, respectively, under the bijective map $\Phi:\  {\cal P}^n_L
\longrightarrow{\cal P}^0_L$,  defined by
$(A,M,X)\longmapsto (A,ML^n,X)$.
So, in the following we simply consider ${\cal P}^0_L$, ${\cal M}^0_L$ and
write  ${\cal M}^0_L$= ${\cal M}_L$, ${\cal P}^0_L$= ${\cal P}_L$.
     
     
    {\bf Definition 3.2} \quad { \it Let $(A,M,X), (B,N,Y)\in {\cal P}_L$.
In ${\cal M}_L$, define a binary operation as follows
\begin{equation}
(A,M) \odot (B,N)=(A\odot B,M\odot N)
\end{equation}
where}
\begin{eqnarray}
A\odot B&=&A_{*}(Y)-B_{*}(X)+[A,B],\\
M\odot N&=&M_{*}(Y)-N_{*}(X)+[M,B]-[N,A].
\end{eqnarray}


    Obviously (3.2) is a skew-symmertric and bilinear operation.


    {\bf Theorem 3.1} \quad { \it Let $(A,M,X), (B,N,Y)\in {\cal P}_L$, then
$(A\odot B,M\odot N,[X,Y])\in {\cal P}_L$, and ${\cal M}_L$
form a Lie algebra under the operation (3.2).}

        {\bf Proof}\quad Since $({\cal V}^N,\ [\cdot,\cdot])$ builds up a Lie
algebra under the usual commutator operation, we have
\begin{eqnarray*}  \left[\left[A,B\right],L\right]&=&\left[\left[L,B\right],A\right]-\left[\left[L,A\right],B\right] \\
             &=&\left[L_{*}(X)-M,B\right]-\left[L_{*}(Y)-N,A\right] \\
            &=&\left[L_{*}(X),B\right] -\left[L_{*}(Y),A\right]
            +\left[N,A\right]-\left[M,B\right]. \end{eqnarray*}
For arbitrary $L\in {\cal V}^N,\ X, Y \in {\cal B}^m$, we also have
$$
(L_{*}(X))_{*}(Y)-(L_{*}(Y))_{*}(X)=L_{*}(\left[X,Y\right]). $$
Thus, {\small
\begin{eqnarray*} \left[A\odot B,L\right]&=&\left[A_{*}(Y)-B_{*}(X)+\left[A,B\right],L\right]\\
               &=&\left[A_{*}(Y),L\right]-\left[B_{*}(X),L\right]+\left[L_{*}(X),B\right]-\left[L_{*}(Y),A\right]+\left[N,A\right]-\left[M,B\right]\\
               &=&(\left[A,L\right])_{*}(Y)-(\left[B,L\right])_{*}(X)+\left[N,A\right]-\left[M,B\right]\\
               &=&(L_{*}(X))_{*}(Y)-(L_{*}(Y))_{*}(X)-M_{*}(Y)+N_{*}(X)+\left[N,A\right]-\left[M,B\right]\\
               &=&L_{*}(\left[X,Y\right])-M\odot N. \end{eqnarray*}}
That means $(A\odot B,M\odot N,[X,Y])\in {\cal P}_L$.
           
           
  Now, we shall prove the Jacobi identity. Choosing
any  $(A_i, M_i, X_i)\in {\cal P}_L$, $i=1,2,3$, then we have
\begin{eqnarray*}(A_1\odot A_2)\odot A_3+c. p.&=&(A_{1*}(X_2)-A_{2*}(X_1)+\left[A_1,A_2\right])\odot A_3+c. p.\\
                               &=&\left[\left[A_1,A_2\right],A_3\right]+c. p.\\
                               &=&0.    \end{eqnarray*}
Similarly, we can show the following equality
$$
(M_1\odot M_2)\odot M_3+c. p.=0. \ \ \ \ \ \ \ \ \ \ \ \ \ \ \ \ \ \ \ \ (*) $$
which completes the proof. \hfill \rule{2mm}{4mm}


    {\bf  Corollary 3.1} \quad { \it  The set of all vector fields
    $V({\cal M}_L)$ forms a Lie subalgebra of ${\cal B}^m$ with
    regard to the operation (1.2).}


    Denote the vector fields of $(A,M)$ and $(B, N)$ by $X$ and $Y$, respectively,
then $u_t=X, \ u_t=Y$ represent the two {\bf different} hierarchies of NLEEs respectively
determined by $M,N$. Theorem 3.1 shows that there is universal algebraic structure
for the {\bf different} hierarchies of NLEEs, and if both $u_t=X$,
and $u_t=Y \quad (X, Y\in {\cal B}^m)$ have GLR,
then so does the new hierarchy of equations $u_t=[X,Y]$ produced by
$X,Y$.


For the given spectral operator $L \in {\cal V}^N$, we now consider
the following adjoint map
\begin{equation}
ad_L: A\longmapsto M=ad_LA=[L,A], \  \forall A \in {\cal V}^N.
\end{equation}
Then according to the original definition of $r$-matrix \cite{5},
we have the following theorem.


    {\bf Theorem 3.2} \quad {\it The adjoint map $ad_L$ is
    an \ma.}


    {\bf Proof}\quad   For any $A, B\in {\cal V}^N$,
write $M=ad_LA, N=ad_LB.$ Then we have
\begin{eqnarray*}\left[A,B\right]_{ad_L}&\stackrel{\triangle}{=}&
 \left[ad_LA,B\right]+\left[A,ad_LB\right]\\
             &=&\left[M,B\right]+\left[A,N\right]\\
             &=&M\odot N.
\end{eqnarray*}
The last equality holds because the associated vector fields are
obviously zero. And Eq. $(*)$ implies that
$[A,B]_{ad_L}$ satisfies the Jacobi identity. Thus the adjoint map
$ad_L$ is an $r$-matrix.
\hfill \rule{2mm}{4mm}


In the last section we shall illustrate that
through giving several examples.


\section{Nonuniversal algebraic structure and \ma}
\setcounter{equation}{0}
For a given spectral operator $L\in {\cal V}^N$ and
integer $n\in Z$, in the above section we discussed the 
Manakov operator pair $(A,M)$, the universal
Lie algebraic structure and $r$-matrix available for
{\bf different} hierarchies of NLEEs.
Now, for a given $N\times N$ matrix operator $M$ and a spectral operator
$L\in {\cal V}^N$, we study the operator algebra and $r$-matrix
which can be attached only to the {\bf underlying} hierarchy of NLEEs.


    Let us first give some conventions in this section:
    (i) $M$ is invertible; (ii) For a given
$L\in {\cal V}^N$, ${\cal V}^N_L$ stands for all matrix operators
$S:{\cal B}^N \rightarrow {\cal B}^N$
possessing the following form
   $S=\sum_{\alpha\in Z}P_{\alpha}(u)L^{\alpha},
P_{\alpha}(u)\in {\cal B}$,
where $\sum_{\alpha\in Z}$ is a finite sum. Next, we introduce
the following definition.
                 
                 
    {\bf Definition 4.1} \quad {\it Let
    $L\in {\cal V}^N$ and $M$  be a spectral operator and an
    $N\times N$ matrix operator, respectively.
If there exist a vector field $X\in {\cal B}^m$
and operators $A,P\in {\cal V}^N_L$
such that
\begin{equation}
[A,L]+MP=L_{*}(X)
\end{equation}
then $(A,P)$ is said to be a $LM$ operator pair of L. The set of all
such pairs is denoted by ${\cal L}^M_L$.
$X$ is called the vector field
of $(A,P)$ associated with
$LM$. The set of all associated vector fields is denoted
by $V({\cal L}^M_L)$.
Furthermore, we denote the set of all triples $(A,P,X)$ by
${\cal R}^M_L$.}


For a given $L\in {\cal V}^N$ and an $N\times N$ matrix operator
$M$ or $\tilde{M}$ theorem 2.1 and equation (2.8)
assure that there exists a triple $(A,P,X)\in{\cal R}^M_L$ satisfying (4.1).
Definition 4.1 directly leads to the following proposition.


    {\bf Proposition 4.1}


    { (i) \it   The vector field associated with
each $LM$ operator pair is unique.


    {\rm (ii)} Both ${\cal L}^M_L$ and ${\cal R}^M_L$ are
linear spaces.}


    If for given operators $L,M$ there exist $A,P\in {\cal V}^N_L$
such that (4.1) holds, then obviously the evolution equation $u_t=X$
has the following representation
(also called generalized Lax representation (GLR))
\begin{equation}
L_t=[A, L]+MP.
\end{equation}
Now, we define a binary operation in ${\cal L}^M_L$.


    {\bf Definition 4.2} \quad
    { \it Let $(A,P),(B,Q)\in {\cal L}^M_L, X, Y\in V({\cal L}^M_L)$
    respectively be the vector fields of $(A,P), (B,Q)$. Declare
    a binary operation
\begin{equation}
(A,P)\ominus (B,Q)=(A\ominus B,P\ominus Q)
\end{equation}
through}
\begin{eqnarray}
A\ominus B&=&A_{*}(Y)-B_{*}(X)+[A,B],\\
P\ominus Q&=&P_{*}(Y)-Q_{*}(X)+[A,Q]-[B,P]\nonumber\\
          & &+M^{-1}(M_{*}(Y)-[B,M])P-M^{-1}(M_{*}(X)-[A,M])Q.
\end{eqnarray}


    {\bf Proposition 4.2}


     { (i) \it Eq. (4.3) is a skew-symmetric,
    bilinear binary operation.


    {\rm (ii)} ${\cal V}^N_L$
is closed under the operations (4.4) and (4.5).}


     {\bf Proof}\quad
     The prove follows directly from Definition 4.2. \hfill \rule{2mm}{4mm}


    {\bf Theorem 4.1}\quad { \it  Let $(A,P,X),(B,Q,Y)\in {\cal R}^M_L$, then
$(A\ominus B,P\ominus Q,[X,Y])\in {\cal R}^M_L$, where $[X,Y]$ is defined
by (1.2). Thus under the operation (4.3)
${\cal L}^M_L$ forms an algebra,
and $(V({\cal L}^M_L), [\cdot,\cdot])$
composes a Lie subalgebra of ${\cal B}^m$.}


     {\bf Proof}\quad Because $(A,P,X),(B,Q,Y)\in {\cal R}^M_L$, and
\begin{eqnarray*}
    \left[\left[A,B\right],L\right]&=&\left[\left[L,B\right],A\right]-\left[\left[L,A\right],B\right] \\
             &=&\left[L_{*}(X),B\right]-\left[L_{*}(Y),A\right]+\left[MQ,A\right]-\left[MP,B\right], \end{eqnarray*}
we have
\begin{eqnarray*}\left[A\ominus B,L\right]&=&\left[A_{*}(Y),L\right]-\left[B_{*}(X),L\right]+\left[\left[A,B\right],L\right] \\
               &=&(L_{*}(X))_{*}(Y)-(L_{*}(Y))_{*}(X)-(MP)_{*}(Y)\\
               & &+(MQ)_{*}(X)+\left[MQ,A\right]-\left[MP,B\right]\\
               &=&L_{*}(\left[X,Y\right])-M(P\ominus N). 
               \end{eqnarray*}
which completes the proof. \hfill \rule{2mm}{4mm}


     For a given spectral operator $L$ and an $N\times N$ matrix operator
$M$, denote the vector fields of $(A,P), (B,Q)$ by $X,Y$ respectively. Then
from section 2 we know $u_t=X, u_t=Y$ are the two different
NLEEs in the {\bf same} hierarchy.
Theorem 4.1 reveals that there exists an algebraic
structure available for all equations in the {\bf same} hierarchy.
And if
$u_t=X,\ u_t=Y$ $(X, Y\in {\cal B}^m)$ have the GLR (4.2), then
the evolution equation $u_t=[X,Y]$ is still in the same hierarchy,
and possesses the GLR (4.2), too.


    {\bf Remark 4.1}\quad
In general, ${\cal L}^M_L$ is not forming a
Lie algebra under the operation (4.3),
because the Jacobi identity can not be guaranteed.
Nevertheless, the subset
$S^M_L\subset {\cal L}^M_L$, considered below,
 is an exception.


    Set $S^M_L=\{(A,P)\in {\cal V}^N_L \times {\cal V}^N_L \vert
    \  P=M^{-1}ad_LA\}$,
then $S^M_L$ is corresponding to the stationary system $X(u)=0$ of
evolution equation $u_t=X(u)$.
          
          
    {\bf Theorem 4.2} \quad { \it For all $ (A,P)\in S^M_L$, define a map
$r^M: A \longmapsto P=M^{-1}ad_LA$.
The map $r^M$ is an \ma  under the operation (4.5) iff
$M=aI,\ a\not=0, a\in R$.}
          
          
    {\bf Proof}\quad  For any $(A,P),(B,Q)\in S^M_L$, define
$$\left[A,B\right]_{r^M}\stackrel{\triangle}{=}\left[r^M(A),B\right]+\left[A,r^M(B)\right].$$
Then
$$\left[A,B\right]_{r^M}=\left[P,B\right]+\left[A,Q\right]=P\ominus Q \Longleftrightarrow
M=aI,\ a\not=0, \ a\in R,$$
i.e. the map $r^M$ is an $r$-matrix $\Longleftrightarrow M
=aI,\ a\not=0, \ a\in R.$
\hfill \rule{2mm}{4mm}

$ $

Since $M$ and $\tilde{M}$ can be fixed arbitrarily
we have found two algebraic operator structures,
namely a universal one being independent of the hierarchy in the category and
a nonuniversal one depending on the underlying hierarchy.
In addition, in this procedure we have found two kinds of adjoint
maps being $r$-matrices.

The two algebraic structures are associated with the category of NLEEs
(2.11) which includes both the integrable and the non-integrable cases
(see Remark 2.4). Therefore, here our algebraic structures are not
contained in any $W$-algebras which is usually suitable for
the integrable hierarchy such as the KP etc.

In the next section, we shall give two reductions of the algebraic structure
and the related $r$-matrix.

\section{Two reductions: Virasoro algebra and \ma of isospectral and
non-isospectral Lax operator}
\setcounter{equation}{0}


If we choose $M=0$ in Definition 4.1,
then we have $[A,L]=L_{*}(X)$.
That means  $A$ is an isospectral ($\lambda_t=0$) Lax operator.
Set $[B,L]=L_{*}(Y)$, then the operation
$A\ominus B$ defined by (4.4) forms
an algebraic structure of the isospectral Lax operator,
which just coincides with the result described in Ref. \cite{26}.
In this case, the $r$-matrix is zero, i.e.
$ad_LA=0, \ \forall A\in {\cal V}^N.$


    In this section, we always choose $M=\tilde{M}=I$
and assume
that the conditions of Theorem 2.1 hold. Then, by Theorem 2.1, we obtain
$$
(W_m,L^{m\eta},\sigma_m)\in {\cal R}_L^{I}, \  m\in Z,$$
where $W_m$ is expressed through (2.10),
$\sigma_m$ stands for the corresponding vector field.
Therefore $W_m$ is a sequence of non-isospectral
($\la_t=\la^{m\eta},m\in Z$) Lax operators and
this matches with choosing $A=W_m,\ P=L^{m\eta}$
($m\in Z$),\ $X=\sigma_m$ in (4.1).
By Theorem 4.1 $\{(W_m,L^{m\eta})$,$m\in Z\}$ represents an algebra under the operation
(4.3), which is called the non-isospectral Lax operator algebra
of the spectral operator $L$.
In the stationary case where
$\sigma_i=\sigma_j=0$ the following holds.


    {\bf Theorem 5.1}\quad { \it
A realization of the operations (4.5) and (4.4)
on pairs 
$(W_i,L^{i\eta}),\\ (W_j,L^{j\eta})\in S^{I}_L$
is given by
\begin{equation}
L^{i\eta}\ominus L^{j\eta}=(\vert i \vert - \vert j \vert)
L^{(i+j-1)\eta}, \quad \forall i,j \in Z,
\end{equation}
\begin{equation}
W_i\ominus W_j=(\vert i \vert - \vert j \vert)W_{i+j-1}, \quad \forall i,j\in Z,
\end{equation}
respectively.
}
                                   
                                   
     {\bf Proof}\quad For $(W_i,L^{i\eta}), (W_j,L^{j\eta})\in S^{I}_L$,
we have
$$\left[W_i,L\right]=-L^{i\eta}, \quad \left[W_j,L\right]=-L^{j\eta}.$$
Thus, in the case $i,j\geq 0$,
\begin{eqnarray*}L^{i\eta}\ominus L^{j\eta}&=&\left[L^{i\eta},W_j\right]-\left[L^{j\eta},W_i\right]\\
                           &=&\sum^{i-1}_{k=0}L^{(i-1-k)\eta}IL^{(k+j)\eta}
                           -\sum^{j-1}_{k=0}L^{(j-1-k)\eta}IL^{(k+i)\eta}\\
                           &=&iL^{(i+j-1)\eta}-jL^{(i+j-1)\eta}\\
                           &=&(i-j)L^{(i+j-1)\eta}.\end{eqnarray*}
Similarly, Eq. (5.1) holds for the other three cases
$i\geq 0,j\leq 0;\ i\leq 0,j\leq 0;\ i\leq 0,j\geq 0$.


Eq. (5.2) can be directly obtained by (5.1) and Theorem 4.1. \hfill \rule{2mm}{4mm}


    {\bf Corollary 5.1}\quad If $M=\tilde{M}=I$, under the
operation (5.1)  the map $r^{I}: W_i\longmapsto L^{i\eta}$
is an $r$-matrix.


    {\bf Proof}\quad  This can be directly derived from Theorem 5.1 and Theorem 4.3. \hfill \rule{2mm}{4mm}


    {\bf Remark 5.1}\quad  Theorem 5.1 and Corollary 5.1 actually
describe the Lie algebraic structure of the Lax operator for the stationary
equation $\sigma_j=0$ ($j\in Z$) and the $r$-matrix of a
concrete form of an operation (4.5) and (4.4), respectively.


    For the usual non-stationary vector field $\sigma_j\not= 0$ ($j\in Z$)
in the non-isospectral case, (5.1) and (5.2) do not hold.
But, we have the following results.


    {\bf Theorem 5.2}\quad { \it
Let $(W_j,L^{j\eta})\in {\cal L}^{I}_L$,
$j\in Z$, then for any $i,j\in Z$, $L$ satisfies
the relation:
\begin{equation}
L^{i\eta}\ominus L^{j\eta}=(\vert i \vert - \vert j \vert)
\eta L^{(i+j+1)\eta-1}, \quad \forall i,j \in Z.
\end{equation}}


     {\bf Proof}\quad We give the proof only
     for the case of $i\geq0,\ j\geq0$. The other cases
     are shown analogously.


Let $ (W_i,L^{i\eta},\sigma_i),(W_j,L^{j\eta},\sigma_j)\in {\cal R}_L^{I}$,
then we have
\begin{eqnarray*}
(L^{i\eta})_{*}(\sigma_j)&=&\sum^{i-1}_{k=0}L^{(i-1-k)\eta}
  L^{\eta}_{*}(\sigma_j)L^{k\eta}\\
                        &=&\sum^{i-1}_{k=0}L^{(i-1-k)\eta}([W_j,
                        L^{\eta}]+\eta L^{(j+1)\eta+\eta-1})L^{k\eta}\\
                        &=&[W_j,L^{i\eta}]+i\eta L^{(i+j+1)\eta-1},
\end{eqnarray*}
and
$$(L^{j\eta})_{*}(\sigma_i)=[W_i,L^{j\eta}]+j\eta L^{(i+j+1)\eta-1}.$$
So, by Eq. (4.5) and noticing $M=I$, we obtain
\begin{equation}
L^{i\eta}\ominus L^{j\eta}=(i-j)\eta L^{(i+j+1)\eta-1},
\quad \forall i,j \in Z^{+},
\end{equation}
which is the desired result.
 \hfill \rule{2mm}{4mm}


Eqs. (5.1), (5.2) and (5.3) are three special Virasoro algebras,
 namely, without
central extension. Because here we do calculations based
on our definitions of binary operations (4.4) and (4.5),
they have no central extensions.

    {\bf Remark 5.2}\quad
For the usual non-stationary vector field
$\sigma_j\not= 0$ ($j\in Z$) in the non-isospectral case
the operation (5.1) does not always
satisfy the Jacobi identity, (see Remark 4.1).
Thus Corollary 5.1 does not hold in general.
                            
                            
    {\bf Remark 5.3}\quad A particular case of Theorem 5.2 is:
    $\eta=1.$ Then Eq. (5.3) becomes
\begin{equation}
L^{i}\ominus L^{j}=(\vert i \vert - \vert j \vert)
 L^{i+j}, \quad \forall i,j \in Z,
\end{equation}
which implies the following Eqs.
\begin{equation}
W_{i}\ominus W_{j}=(\vert i \vert - \vert j \vert)
 W_{i+j}, \quad \forall i,j \in Z,
\end{equation}
and
\begin{equation}
[\sigma_i,\sigma_j]=(\vert i \vert - \vert j \vert)
 \sigma_{i+j}, \quad \forall i,j \in Z.
\end{equation}
Theorem 5.2 reveals that under Eq. (5.5) for the same non-isospectral hierarchy 
the following holds:
if $u_t=\sigma_m$ and $\ u_t=\sigma_n$ respectively possess
the non-isospectral Lax operators
$W_m$ and $\ W_n$, then $u_t=\sigma_{m+n}$ still possesses the non-isospectral
Lax operator $\frac{1}{\vert m \vert - \vert n \vert}W_m\ominus W_n, \  \forall m,n \in Z$.
Thus, the Virasoro operator algebras (without the central extension) for the
non-isospectral hierarchy of NLEEs
is reflected by Eqs. (5.5)-(5.7).

   {\bf Remark 5.4}\quad  If we choose $M=0$ and $M=I$
respectively, then under the algebraic operation (3.3) we can also 
have  the Virasoro algebra of the Lax
operator for the isospectral hierarchy and the non-isospectral hierarchy,
which is actually a special case of universal algebraic structure.

            
\section{\bf Some examples}
\setcounter{equation}{0}
Through taking several examples, we illustrate
our methods. For our convenience, we make the following conventions:
\begin{eqnarray*}
f^{(m)}=\left\{\begin{array}{ll}
               \frac{\pa^m}{\pa x^m}f=f_{mx}, & m\ge0,\\
               \underbrace{\int\ldots\int}_{-m}f dx, & m<0,
               \end{array}
         \right.\ \ \  \sum=\left\{\begin{array}{ll}
   \sum_{j=0}^{m-1}, & m>0, \\
   0, & m=0,\\
   -\sum_{j=m}^{-1}, & m<0, \end{array}\right.
\end{eqnarray*}
$f_t=\frac{\pa f}{\pa t}$, $f_{mxt}=\frac{\pa^{m+1} f}{\pa t\pa x^m}\
(m\geq 0)$,
$\pa=\frac{\pa}{\pa x}$, $\pa^{-1}$ is the inverse of $\pa$, i.e.
$\pa \pa^{-1}=\pa^{-1}\pa=1$,
$\pa^mf$ means the operator $\pa^mf$ acts on some function $g$, 
i.e. $\pa^mf\cdot g=\pa^m(fg),\ m\in Z$.
$C^k_m$ stands for the combinatorial
constants: $C_m^k=\frac{m(m-1)\cdots(m-k+1)}{k!}$, $i$ an
imaginary unit satisfying $i^2=-1$, and $I_{2\times2}$ the $2\times2$
unit matrix.


In the spectral problems
(\ref{6.1}), (\ref{6.35}) and (\ref{l1})
the function $u$ stands for the potential function, and the potential
functions in spectral problems (\ref{6.341}) and (\ref{6.4.1})
are denoted by $q$, $r$. In those spectral problems, $\la$ is
always assumed to be a spectral parameter.
The domain of the spatial variable
$x$ is $\Omega$ which becomes equal to
$(-\infty,\ +\infty)$ or $(0, \ T)$,
while the domain of the time variable $t$
is the positive time axis $R^{+}=\{t|t\in R, \ t\geq 0\}$.
In the case $\Omega=(-\infty,\ +\infty )$
the decaying condition at infinity
and in the case $\Omega=(0,\ T)$
the periodicity condition for the potential function,
is imposed.


$ $


{\large \bf 6.1}  \quad  Consider the Burgers spectral problem \cite{27}
\begin{equation}
L\cdot y=\la y, \ L=L(u)=\pa +u. \label{6.1}
\end{equation}
Choosing the recursion operator
${\cal L}= \pa+\pa u \pa^{-1}$
leads to
\be
  {\cal L}\cdot y_x = \la y_x. \label{6.2}
\ee
Obviously,
$L_{*}(\xi)=\xi, \forall \xi\in {\cal B}$
i.e. $L_{*}$ is an identity operator. In this case,
the Lenard's operators pair is chosen as $J=1$, and $K={\cal L}$.


The Lenard recursive sequence $\{G_{j}\}_{j=-\infty}^{\infty}$
($G_j={\cal L}^j\cdot M, \ j\in Z$) gives the Burgers category of
NLEEs
\beq
u_t=
{\cal L}^m\cdot M=\left(e^{-u^{(-1)}}(e^{u^{(-1)}}M^{(-1)})^{(m)}\right)_x,
\ m\in Z, \label{6.3}
\eeq
where $M\in {\cal B}$ is an arbitrarily given function,
and ${\cal L}=\pa e^{-u^{(-1)}}\pa e^{u^{(-1)}}\pa^{-1}$,
${\cal L}^{-1}=\pa e^{-u^{(-1)}}\pa^{-1}e^{u^{(-1)}}\pa^{-1}$
which implies ${\cal L}^j=\pa e^{-u^{(-1)}}\pa^j e^{u^{(-1)}}\pa^{-1}$,
$j\in Z$.


    For an arbitrary $G\in {\cal B}$, the operator equation
$  [V, L]=L_{*}({\cal L}\cdot G) - L_{*}(G)L$,
which matches with choosing $\beta=0,\alpha=1$ in (2.8),
has the following solution
\be
  V=V(G)=-G+G^{(-1)} \pa. \label{6.4}
\ee
Thus the category (\ref{6.3})
possesses the generalized Lax representation (GLR)
\beq
  L_{t}=\left[W_{m}, L\right] + ML^m, \ m\in Z, \label{6.5}
\eeq
with $W_{m}=M^{(-1)}L^m-L^m\cdot M^{(-1)}$,
$L^m=e^{-u^{(-1)}}\pa^m e^{u^{(-1)}},\
m\in Z$.


The transformation $u=(\ln v)_x$ yields a simple form of Eq. (\ref{6.3}):
\be v_t=(vM^{(-1)})^{(m)},\ \ m\in Z, \label{6.6} \ee
which has the GLR $L_t=[W_m,L]+ML^m$ with $L=\pa+(\ln v)_x,$ $\
W_m=v^{-1}\left(M^{(-1)}\pa^mv-\right.\\ \left.(vM^{(-1)})^{(m)}\right)$.


Apparently, if $M$ is chosen to be independent of $v \ (v=e^{u^{(-1)}})$,
then the category (\ref{6.3}) is linearized.
Thus, (\ref{6.3}) includes many linearized hierarchies. Now, let us discuss
reductions of the category (\ref{6.3}) or (\ref{6.6}).


$ $


{\bf I. Positive Case $(m=0,1,2,\ldots)$}


 In this case, the Lax operator $W_m$ can be written as
\be
W_m=v^{-1}\sum_{k=1}^mC_m^kv^{(m-k)}\left(M^{(-1)}\pa^k-M^{(k-1)}\right).
\label{6.7}
\ee
\begin{itemize}
 \item With $M=0$, $0^{(-1)}=1,$ the positive order category of (\ref{6.3})
   reads as the well-known Burgers hierarchy
   \be u_t=\left((\pa+u)^m\cdot 1\right)_x. \label{6.8} \ee
\end{itemize}
Particularly, with $m=2$ it becomes the Burgers equation
$u_t=u_{xx}+2uu_x$ whose Lax operator is $W_2=\pa^2+2u\pa$
in the standard Lax representation $L_t=[W_2,L]$. This corresponds
to the isospectral case: $\la_t=0$. According to Eq. (\ref{6.6}),
a simple but quite interesting fact is that under the transformation $u=
\frac{v_x}{v}$ the whole Burgers hierarchy (\ref{6.8}) is linearized as
\be v_t=v_{mx}, \ m=0,1,2,\cdots.  \label{6.9} \ee
Eq. (\ref{6.9}) can be solved very easily and have the standard Lax pair
$W_m=v^{-1}\sum_{k=1}^m
C_m\\^kv^{(m-k)}\pa^k $ and $L=\pa+\frac{v_x}{v}$.
In this way, the solutions of all equations in the Burgers hierarchy (\ref{6.8})
can be worked out.


\begin{itemize}
 \item With $M=a$, $a^{(-1)}=ax+f(t),\ a\in R, \ f(t)\in C^{\infty}(R)$,
 the positive order category of (\ref{6.3})
becomes the non-isospectral ($\la_t= a\la^m$) Burgers hierarchy
   \be u_t=\left((\pa+u)^m\cdot (ax+f(t))\right)_x. \label{6.10} \ee
\end{itemize}
A representative equation $(m=2)$ of Eq. (\ref{6.10}) is
\be u_t=\left(ax+f(t)\right)(u_{xx}+2uu_x)+3au_x+au^2 \label{6.11} \ee
possessing the GLR $L_t=[W_2,L]+aL^2$ with $W_2=(ax+f(t))(\pa^2+2u\pa)-2au$
and $L=\pa+u$. By virtue of $M=a$ and $u=(\ln v)_x$
Eq. (\ref{6.10}) is linearized as
\be
v_t=\left(ax+f(t)\right)v_{mx}+mav_{(m-1)x} \label{6.12}\ee
which can be solved.
Eq. (\ref{6.12}) has the generalized Lax operator (GLO)
$W_m=(ax+f(t))v^{-1}\sum_{k=1}^mC_m^kv^{(m-k)}\pa^k-mav^{-1}v^{(m-1)}$.
Particularly, Eq. (\ref{6.11}) has a linearization equation ($m=2$)
\be v_t=\left(ax+f(t)\right)v_{xx}+2av \label{6.13} \ee
possessing the GLO $W_2=(ax+f(t))(\pa^2+2v^{-1}v_x\pa)-2av^{-1}v_x$.
In a general case, $M$ can be extended as $M=\sum_{j=0}^nc_j(t)x^j,
\ c_j(t)\in C^{\infty}(R)$, which will be considered below.


\begin{itemize}
 \item With $M=\sum_{j=0}^nc_j(t)x^j, \ c_j(t)\in C^{\infty}(R)$,
 the positive order category of (\ref{6.3})
reads as a non-isospectral ($\la_t= \left(\sum_{j=0}^nc_j(t)x^j\right)\la^m$)
hierarchy
   \be u_t=\left((\pa+u)^m\cdot \left(f(t)+\sum_{j=0}^nc_j(t)\frac{x^{j+1}}
   {j+1}\right)\right)_x
    \label{6.14} \ee
\end{itemize}
where an arbitrary $f(t)\in C^{\infty}(R)$ is attached by virtue
of integration with respect to $x$.
Of course, Eq. (\ref{6.14})
is easily linearized as
\beq
v_t=\frac{\pa^m}{\pa x^m}\left(vf(t)+v
\sum_{j=0}^n\frac{c_j(t)}{j+1}x^{j+1}\right) \label{vtt}  
\eeq
via $u=(\ln v)_x$. Eq. (\ref{vtt}) has the generalized Lax operator
\beqq
W_m=v^{-1}\sum_{k=1}^mC_m^kv^{(m-k)}\left(M^{(-1)}\pa^k-M^{(k-1)}\right)
\eeqq 
with
$$M^{(-1)}=f(t)+
\sum_{j=0}^n\frac{c_j(t)}{j+1}x^{j+1}.$$

\begin{itemize}
 \item With $M=(u^{-1})_x$, $M^{(-1)}=u^{-1},$
 the positive order category of (\ref{6.3})
reads the following hierarchy of NLEEs:
   \be u_t=\left((\pa+u)^m\cdot u^{-1}\right)_x, \ m=0,1,2\ldots. \label{6.19} \ee
\end{itemize}
A representative equation of (\ref{6.19}) is
\be u_t=\left(\frac{1}{u}\right)_{xx} \label{6.20} \ee
with the GLO $W_0=-(u^{-1})_x+u^{-1}\pa$.


$ $


{\bf II. Negative Case $(m=-1,-2,\ldots)$}
\begin{itemize}
 \item With $M=0$, the generator $G_{-1}=\pa
 e^{-u^{(-1)}}\pa^{-1}e^{u^{(-1)}}\pa^{-1}\cdot0$
is determined by the following two seed functions
 \end{itemize}
\be \bar{G}_{-1}=f(t)(e^{-u^{(-1)}})_x \label{6.21} \ee
and
\be \tilde{G}_{-1}=g(t)\left(e^{-u^{(-1)}}(e^{u^{(-1)}})^{(-1)}\right)_x
      \label{6.22} \ee
where $f(t),\ g(t)\in C^{\infty}(R)$ are two arbitrarily given functions.
Apparently, the seed function (\ref{6.21}) produces the following
isospectral ($\la_t=0$) negative order hierarchy of (\ref{6.3})
   \be u_t=f(t)\left(e^{-u^{(-1)}}1^{(m)}\right)_x,
   \ m<0, \ m\in Z, \label{6.23} \ee
i.e.
   \be u_t=f(t)e^{-u^{(-1)}}\sum_{k=0}^{-m-1}c_k
   \frac{x^{-m-k-2}(-m-k-1-xu)}{(-m-k-1)!}, \ c_0=1,
   \label{6.24} \ee
where $c_k=c_k(t)\in C^{\infty}(R) \ (-m-1\ge k\ge 1)$
is arbitrarily given.
Thus although Eq. (\ref{6.23}) is nonlinear, we have its general solution:
\be u(x,t)=\frac{\sum_{k=0}^{-m-2}c_k(t)
   \frac{x^{-m-k-2}}{(-m-k-2)!}\pa_t^{-1}f(t)+h'(x)}
{\sum_{k=0}^{-m-1}c_k(t)
   \frac{x^{-m-k-1}}{(-m-k-1)!}\pa_t^{-1}f(t)+h(x)}, \ \forall h(x), \ c_k(t)\in C^{\infty}(R),
   \label{6.25} \ee
where $\pa_t^{-1}f(t)=\int f(t) dt,\ c_0(t)=1, \ h'(x)=
\frac{d}{dx}h(x)$. Of course, Eq. (\ref{6.24}) has the standard Lax
representation $L_t=[W_m,L]$ with $W_m=
-f(t)e^{-u^{(-1)}}\sum_{k=0}^{-m-1}c_k(t)\\
   \frac{x^{-m-k-1}}{(-m-k-1)!}$.


On the other hand, the seed function (\ref{6.22})
generates the following isospectral ($\la_t=0$) negative order
hierarchy of (\ref{6.3})
\be u_t=g(t)\left(e^{-u^{(-1)}}\left(e^{u^{(-1)}}\right)^{(m)}
\right)_x,
   \ m<0, \ m\in Z, \label{6.26} \ee
which is a hierarchy of integro-differential equations and
can be changed to the linear differential equations
\be v_{-mxt}=g(t)v,
\ m<0,\ m \in Z, \label{6.27}\ee
via the transformation $u=v^{-1}v_x$. The Lax operator
$W_m$ of (\ref{6.26}) or (\ref{6.27}) is $W_m=-g(t)
e^{-u^{(-1)}}(e^{u^{(-1)}})^{(m)}$ \ or \ $W_m=-g(t)
v^{-1}v^{(m)}$, $m<0$.


\begin{itemize}
 \item With $M=a$, $a^{(-1)}=ax+f(t),\ a\in R, \ f(t)\in C^{\infty}(R)$,
 the negative order category of (\ref{6.3}) through setting $u=v^{-1}v_x$
reads as the linear equations 
\end{itemize}
\be v_{-mxt}=\left(ax+f(t)\right)v,
\ m<0,\ m \in Z, \label{6.28}\ee
which corresponds to the nonisospectral case: $\la_t=a\la^m$,
and has the GLO $W_m=v^{-1}\left(ax+f(t)\right)\pa^mv-v^{-1}
\left(v(ax+f(t))\right)^{(m)}, \ m<0$. For a general case, we have the following.


\begin{itemize}
 \item Setting $M=\sum_{j=0}^nc_j(t)x^j \ (c_j(t)\in C^{\infty}(R))$
yields a negative order hierarchy of (\ref{6.3})
   \be u_t=\left(e^{-u^{(-1)}}\pa^m e^{u^{(-1)}}\cdot
   \sum_{j=0}^nc_j(t)\frac{x^{j+1}}{j+1}\right)_x,
   \ m<0, \ m\in Z, \label{6.29} \ee
\end{itemize}
which corresponds to the nonisospectral case
$\la_t= \left(\sum_{j=0}^nc_j(t)x^j\right)\la^m$, and can be
linearized as
\be v_{-mxt}=v\sum_{j=0}^nc_j(t)\frac{x^{j+1}}{j+1},
\ m<0,\ m \in Z, \label{6.30}\ee
via $u=v^{-1}v_x$. Eq. (\ref{6.30}) has the Lax operator
 $$W_m=v^{-1}\sum_{j=0}^n\frac{c_j(t)}{j+1}\left(x^{j+1}\pa^mv-
(vx^{j+1})^{(m)}\right),\ \ m<0. $$
\begin{itemize}
 \item With $M=(\frac{v}{v_x})_x$, $\pa^{-1}M=\frac{v}{v_x}$,
  the associated negative order hierarchy of (\ref{6.3}) is
\end{itemize}
\be v_{-mxt}=\frac{v^2}{v_x}, \
\ m<0,\ \ m\in Z, \label{6.31}\ee
which has a representative equation
($m=-1$)
\be v_xv_{xt}=v^2 \label{6.32}\ee
with the Lax operator $W_{-1}=\frac{1}{v_x}\pa^{-1}v
-\frac{1}{v}(\frac{v^2}{v_x})^{(-1)}$.


$ $


Through choosing different $M$, we still have other hierarchies of
(\ref{6.3}). Because of the arbitrariness of $M$, all results
in sections 3 -- 5 are valid for the Burgers (B) spectral problem
(\ref{6.1}). Particularly, the $r$-matrix $ad_L$ becomes
\be ad_L^B: W_m \longmapsto ML^m, \  m\in Z, \label{6.33}
     \ee
where $W_m=M^{(-1)}L^m-L^m\cdot M^{(-1)}, \ L^m=
e^{-u^{(-1)}}\pa^me^{u^{(-1)}}, \ M\in {\cal B}$ is
an arbitrarily given function. And the
$r$-matrix $r^M$ ($M=a\not=0, a\in R$) reads
\be r_B^a: W_m \longmapsto L^m, \  m\in Z, \label{6.34}
     \ee
where $W_m=(ax+f(t))L^m-L^m\cdot(ax+f(t))$. Eqs. (\ref{6.33}) and
(\ref{6.34}) generate the stationary B-categorical systems
$(L^m\cdot M^{(-1)})_x=0$ and $\left(L^m\cdot(ax+f(t))
\right)_x=0$, respectively.

We can also apply the above procedure to other spectral problems.
Now, we list some main results as follows.

$ $

{\large \bf 6.2}  \quad  KdV case. The KdV-Schr\"{o}dinger
spectral problem \cite{29}
\begin{equation}
L\cdot y=\la y, \ L=L(u)=\pa^2 +u,  \label{6.35}
\end{equation}
has the following Lenard operator pair
\be K=\frac{1}{4}\pa^3+\frac{1}{2}(\pa u+u\pa), \ \  J=\pa. \label{6.37}
    \ee
Apparently, $L_{*}(\xi)=\xi, \forall \xi\in {\cal B}$.
Setting $u=-\frac{\phi_{xx}}{\phi}$ yields the product-form of $K$
and its inverse
\beq
K&=&\frac{1}{4}\phi^{-2}\pa\phi^{2}\pa\phi^{2}\pa\phi^{-2},\nonumber\\
K^{-1}&=&4\phi^{2}\pa^{-1}\phi^{-2}\pa^{-1}\phi^{-2}\pa^{-1}\phi^{2}.
         \label{6.38}
         \eeq

Let $M, \ \tilde{M}\in {\cal B}$ be two arbitrarily given functions.
Then the positive order and negative order generators
\be G_0=M^{(-1)}; \ G_{-1}=K^{-1}\cdot\tilde{M}=
4\phi^{2}\pa^{-1}\phi^{-2}\pa^{-1}\phi^{-2}\pa^{-1}\cdot(\phi^{2}\tilde{M}),
        \label{6.39} \ee
leads to the KdV category of NLEEs
\be
   u_t=J\cdot G_m, \ m\in Z, \ G_m=
   \left\{\begin{array}{ll}
   {\cal L}^{m}\cdot G_0, & m\geq0,\\
   {\cal L}^{m+1}\cdot G_{-1}, & m<0,
\end{array}\right. \label{6.40}
\ee
where the recursion operator ${\cal L}$ is given by
$$
{\cal L}=J^{-1}K=\frac{1}{4}\pa^2+\frac{1}{2}(u+\pa^{-1}u\pa)=
\frac{1}{4}\pa^{-1}\phi^{-2}\pa\phi^{2}\pa\phi^{2}\pa\phi^{-2}$$
and its inverse is
$${\cal L}^{-1}=
4\phi^{2}\pa^{-1}\phi^{-2}\pa^{-1}\phi^{-2}\pa^{-1}\phi^{2}\pa.$$


    For an arbitrary $G\in {\cal B}$, the operator equation
$[V, L]=L_{*}(K\cdot G) - L_{*}(J\cdot G)L$ has the
following operator solution
\be
  V=V(G)=-\frac{1}{4}G_x+\frac{1}{2}G\pa \label{6.41}
\ee
which implies that
the KdV category (\ref{6.40})
possesses the GLR
\be
L_t=\left[W_{m}, L\right] + \bar{M}L^m, \ m\in Z, \ \ \bar{M}=
\left\{\begin{array}{ll}
 M &  m\geq0, \\
 \tilde{M}, & m<0,
\end{array}\right. \label{6.42}
\ee
with the GLO
\be
W_{m}=\sum V(G_{j})L^{m-j-1}.
   \label{6.43}
\ee
Here $V(G_j)$ is determined by (\ref{6.41}) with $G=G_j={\cal L}^j\cdot G_0, \
j\ge0$ or $G=G_j={\cal L}^{j+1}\cdot G_{-1}, \ j<0$, $L=\pa^2+u=
\phi^{-1}\pa\phi^{-2}\pa\phi^{-2}$, and $L^{-1}=
\phi^2\pa^{-1}\phi^2\pa^{-1}\phi$.

In particular, we concern the following reduction. 
\begin{itemize}
 \item With $M=4\left(u^{-\frac{1}{2}}\right)_x$,
 $G_0=M^{(-1)}=4u^{-\frac{1}{2}},$
 the positive order category of (\ref{6.40})
reads as the well-known Harry-Dym hierarchy
   \be u_t=J{\cal L}^m\cdot 4u^{-\frac{1}{2}}, \ m=0,1,2\ldots. \label{6.53} \ee
\end{itemize}
With $m=1$ Eq. (\ref{6.53}) yields the Harry-Dym equation
\be u_t=\left(\frac{1}{\sqrt{u}}\right)_{xxx} \label{6.54} \ee
which has now the GLR $L_t=[W_0,L]+4(u^{-\frac{1}{2}})_xL$ with
$W_0=-(u^{-\frac{1}{2}})_x+2u^{-\frac{1}{2}}\pa$, and apparently
belongs to the KdV category (\ref{6.40});
with $m=2$ Eq. (\ref{6.53}) yields a higher order Harry-Dym equation
\be u_t=\frac{1}{4}\left(\frac{1}{\sqrt{u}}\right)_{4x}
     +u\left(\frac{1}{\sqrt{u}}\right)_{3x}
     +\frac{1}{2}u_x\left(\frac{1}{\sqrt{u}}\right)_{xx}
\label{6.55} \ee
possessing the GLO
$W_1=2u^{-\frac{1}{2}}\pa^3-(u^{-\frac{1}{2}})_x\pa^2+\frac{1}{2}
\left((u^{-\frac{1}{2}})_{xx}+4u^{-\frac{1}{2}}\right)\pa-
\frac{1}{4}(u^{-\frac{1}{2}})_{xxx}+u^{-\frac{1}{2}}u_x.$

So, we have obtained an interesting fact: {\bf the Harry-Dym equation
(\ref{6.54}) can be included in the KdV category (\ref{6.40})
with the generalized Lax operator.} Similar to the process of the
Burgers'
case, we can also have many reduced hierarchies both positive and
negative from Eq. (\ref{6.40}).

$ $


{\large \bf 6.3}  \quad  
AKNS case. The ZS-AKNS spectral problem \cite{30,31}
\begin{equation}
L\cdot y=\la y, \ L=L(q,r)=i\left(\begin{array}{ll}
                           \pa & -q\\
                           r &-\pa
                       \end{array}\right),  \ \ y=
\left(\begin{array}{l}
                     y_1\\
                     y_2
                       \end{array}\right),   \label{6.341}
\end{equation}
has its  Lenard's operators pair
\be 
        K=\left(\begin{array}{lr}
                           q\pa^{-1}q & \frac{1}{2}\pa-q\pa^{-1}r\\
                           \frac{1}{2}\pa-r\pa^{-1}q & r\pa^{-1}r
                       \end{array}\right),  \ \
       J=i\left(\begin{array}{lr}
                           0 & -1\\
                           1 & 0
                       \end{array}\right). \label{6.343}
\ee
Apparently, 
\be
L_{*}(\xi)=\left(\begin{array}{lr}
                           0 & -i\xi_1\\
                           i\xi_2 & 0
                       \end{array}\right), \ \ \xi=(\xi_1,\xi_2)^T\in{\cal B}^2,
                        \label{6.344}
\ee
is an injective homomorphism.


Eq. (\ref{6.343}) gives the recursion operator
\be
{\cal L}=J^{-1}K=\frac{1}{2}i
        \left(\begin{array}{lr}
                           -\pa+2r\pa^{-1}q & -2r\pa^{-1}r\\
                           2q\pa^{-1}q & \pa-2q\pa^{-1}r
                       \end{array}\right).
              \label{6.345} \ee
Choosing two functions
$\theta, \ \sigma \in C^{\infty}(R)$
satisfying 
$\theta_x=\frac{1}{2}\theta^2+r^{-1}r_x\theta-2qr$, \ 
$\sigma_x=\frac{1}{2}\sigma^2+q^{-1}q_x\sigma-2qr$,
leads to  the inverse of ${\cal L}$
\be
{\cal L}^{-1}=K^{-1}J=-2i
        \left(\begin{array}{lr}
    -{\cal E}\left(\pa r^{-1}\pa r^{-1}-2qr ^{-1}\right) & -2{\cal E}\\
       2{\cal F} & {\cal F}\left(\pa q^{-1}\pa q^{-1}-2rq^{-1}\right) 
                       \end{array}\right)
              \label{6.346} \ee
where ${\cal E}, \ {\cal F}$ denote the following two operators
\be
{\cal E}=e^{-\theta^{(-1)}}\pa^{-1}
  e^{\theta^{(-1)}}r\pa^{-1}re^{\theta^{(-1)}}\pa^{-1}e^{-\theta^{(-1)}}, \
{\cal F}=e^{-\sigma^{(-1)}}\pa^{-1}
  e^{\sigma^{(-1)}}q\pa^{-1}qe^{\sigma^{(-1)}}\pa^{-1}e^{-\sigma^{(-1)}}.
  \label{6.347}
\ee

    Let $A, B, C, D\in {\cal B}$ be four arbitrarily given functions, then iff
\begin{equation}
M=\left(\begin{array}{cc}
0 & -B \\
-A & 0
\end{array}\right), \quad \tilde{M}=\left(\begin{array}{cc}
0 & -D \\
-C & 0
\end{array}\right),  \label{6.348}
\end{equation}
the operator equations $ L_{*}(J\cdot G_0)=M$, $L_{*}(K\cdot G_{-1})=\tilde{M}$
give the positive order and negative order generators (function vectors)
\begin{equation}
G_0=
\left(\begin{array}{c}
A\\
B
\end{array}\right), \ \ 
G_{-1}=-2i
\left(\begin{array}{c}
-{\cal E}\cdot(\pa r^{-1}\cdot (r^{-1}C)_x-2qr^{-1}C+2D)\\
 {\cal F}\cdot(\pa q^{-1} \cdot(q^{-1}D)_x-2rq^{-1}D+2C)
\end{array}\right), \label{6.349} 
\end{equation}
which directly leads to the AKNS category of NLEEs:
\be
\left(\begin{array}{c}
q\\
r
\end{array}\right)_t=\left\{\begin{array}{ll}
J{\cal L}^m\cdot(A,B)^T, & m=0,1,2\ldots, \\
J{\cal L}^m\cdot(C,D)^T, & m=-1,-2,\ldots,
\end{array}\right.
\label{6.350} 
\ee
where $J,\ {\cal L}$ and ${\cal L}^{-1}$ are defined
by (\ref{6.343}), (\ref{6.345}) and (\ref{6.346}), respectively.


   For an arbitrarily  given $G=(G^{[1]},G^{[2]})^T \in {\cal B}^2$,
    the operator equation 
    $[V,L]=L_{*}(K\cdot G)-L_{*}(J\cdot G)L$ has the solution
\begin{equation}
  V=V(G)=\frac{1}{2}\left(\begin{array}{lr}
-(rG^{[2]}-qG^{[1]})^{(-1)} & G^{[2]}\\
G^{[1]} &  (rG^{[2]}-qG^{[1]})^{(-1)}
\end{array}\right)  \label{6.351}
\end{equation}
which is obviously a function matrix. 
    Thus, the AKNS category (\ref{6.350}) have
    the GLR:
\beq
  L_t &=&\left[W_{m}, L\right] + \bar{M}L^m, \ m\in Z,\label{6.352}\\
           & & \bar{M}=\left\{\begin{array}{ll}
     \left(\begin{array}{cc}
     0 & B\\
     A & 0
     \end{array}\right), & m\ge0, \\
      \left(\begin{array}{ll}
      0 & D\\
      C & 0
      \end{array}\right), & m<0,
      \end{array}\right.\nonumber
\eeq
with  the GLO
\be
W_{m}  = \sum V(G_{j})L^{m-j-1}, \ m\in Z. \label{6.353}
\ee
Here $V(G_j)$ is given by (\ref{6.351}) with 
$G=G_j={\cal L}^j\cdot(A,B)^T,\ j\geq0$ or
${\cal L}^j\cdot(C,D)^T,\ j<0$, $L$ is defined by (\ref{6.341}),
and its inverse $L^{-1}$ is determined by
\be
 L^{-1}=i
        \left(\begin{array}{lr}
                           {\cal S}\pa q^{-1} & -{\cal S}\\
                           -{\cal T} & {\cal T}\pa r^{-1} 
                       \end{array}\right)
              \label{6.354} \ee
with the operators
${\cal S}=e^{-\rho^{(-1)}}\pa^{-1}
  e^{2\rho^{(-1)}}q\pa^{-1}e^{-\rho^{(-1)}}, \
{\cal T}=e^{-\mu^{(-1)}}\pa^{-1}
  e^{2\mu^{(-1)}}r\pa^{-1}e^{-\mu^{(-1)}}$,
where $\rho$ and $\mu$ are two functions satisfying
$\rho_x=\rho^2+q^{-1}q_x\rho-qr, \  
\mu_x=\mu^2+r^{-1}r_x\mu-qr$.

Here, we omit the reductions and the $r$-matrix representation
 of the AKNS category (\ref{6.350}).

$ $

{\large \bf 6.4}  \quad WKI
(Wadati-Konno-Ichikowa) case. The WKI spectral problem \cite{32}
\begin{equation}
L\cdot y=\la y, \ L=L(q,r)=\frac{1}{1-qr}\left(\begin{array}{ll}
                           i & -q\\
                           -r &-i
                       \end{array}\right)\pa, \ \
                       y=\left(\begin{array}{l}
                       y_1\\
                       y_2
                       \end{array}\right),    \label{6.4.1}
\end{equation}
has the following Lenard's operators pair
\beq
        K &=& \frac{1}{2i}\left(\begin{array}{lr}
   -\frac{1}{2}\pa^2 \frac{q}{p}\pa^{-1}\frac{q}{p}\pa^2 & \pa^3+\frac{1}{2}\pa^2 \frac{q}{p}\pa^{-1}\frac{r}{p}\pa^2 \\
    \pa^3+\frac{1}{2}\pa^2 \frac{r}{p}\pa^{-1}\frac{q}{p}\pa^2 & -\frac{1}{2}\pa^2 \frac{r}{p}\pa^{-1}\frac{r}{p}\pa^2
                       \end{array}\right),  \label{k4}\\
       J &=& \left(\begin{array}{lr}
                           0 & -\pa^2\\
                           \pa^2 & 0
                       \end{array}\right), \ \ \ p=\sqrt{1-qr},
                      \label{6.4.3}
\eeq
which yields the recursion operator ${\cal L}=J^{-1}K$
\beq
{\cal L} &=& \frac{1}{2i}\left(\begin{array}{lr}
               \pa+\frac{r}{2p}\pa^{-1}\frac{q}{p}\pa^2 &
                -\frac{r}{2p}\pa^{-1}\frac{r}{p}\pa^2 \\
                \frac{q}{2p}\pa^{-1}\frac{q}{p}\pa^2 &
                -\pa-\frac{q}{2p}\pa^{-1}\frac{r}{p}\pa^2
                       \end{array}\right).  \label{6.4.4}
\eeq


Apparently, the Gateaux derivative operator $L_{*}(\xi)$ of the spectral
operator $L$ in the direction $\xi=(\xi_1,\xi_2)^T\in {\cal B}^2$ is
\be
L_{*}(\xi)=\frac{1}{1-qr}\left(\begin{array}{lr}
                           q\xi_2 & -i\xi_1\\
                           i\xi_2 & r\xi_1
                       \end{array}\right)L
                        \label{6.4.5}
\ee
which is an injective homomorphism.


Through lengthy calculations, one can obtain the invertible operators
of $L,\ J,\ K$ and ${\cal L}$:
\beq
 L^{-1} & = & \left(\begin{array}{lr}
                           -i\pa^{-1} & \pa^{-1}q\\
                           \pa^{-1}r & i\pa^{-1}
                       \end{array}\right),     \label{6.4.6}\\
 J^{-1} & = & \left(\begin{array}{lr}
                           0 & \pa^{-2}\\
                           -\pa^{-2} & 0
                       \end{array}\right),     \label{6.4.7}\\
  K^{-1} & = & 2i\left(\begin{array}{lr}
  \frac{1}{2}\pa^{-1}r\pa^{-1}r\pa^{-1} & \pa^{-3}-\frac{1}{2}\pa^{-1}r\pa^{-1}q\pa^{-1} \\
  \pa^{-3}-\frac{1}{2}\pa^{-1}q\pa^{-1}r\pa^{-1} & \frac{1}{2}\pa^{-1}q\pa^{-1}q\pa^{-1}
                       \end{array}\right),  \label{6.4.8}\\
  {\cal L}^{-1} & = & 2i\left(\begin{array}{lr}
  \pa^{-1}-\frac{1}{2}\pa^{-1}r\pa^{-1}q\pa & -\frac{1}{2}\pa^{-1}r\pa^{-1}r\pa\\
  \frac{1}{2}\pa^{-1}q\pa^{-1}q\pa & -\pa^{-1}+\frac{1}{2}\pa^{-1}q\pa^{-1}r\pa
                         \end{array}\right).  \label{6.4.9}
\eeq


    Let $A,\ B, \ C, \ D$
    be four arbitrarily given $C^{\infty}$-functions,
    then iff
\begin{equation}
M=\frac{1}{1-qr}\left(\begin{array}{lr}
qA & iB \\
iA & -rB
\end{array}\right)L, \quad \tilde{M}=
\frac{1}{1-qr}\left(\begin{array}{lr}
qC & iD \\
iC & -rD
\end{array}\right)L,  \label{6.4.10}
\end{equation}
the operator equations $ L_{*}(J\cdot G_0)=M$, $L_{*}(K\cdot G_{-1})=\tilde{M}$
have the following solutions
\beq
G_0 & = & \left(\begin{array}{l}
A^{(-2)}\\
B^{(-2)}
\end{array}\right), \label{6.4.11}\\
G_{-1} & = &
\left(\begin{array}{l}
 2iC^{(-3)}-i\pa^{-1}r\pa^{-1}\cdot(rD^{(-1)}+qC^{(-1)})\\
-2iD^{(-3)}+i\pa^{-1}q\pa^{-1}\cdot(rD^{(-1)}+qC^{(-1)})
\end{array}\right), \label{6.4.12}
\eeq
which directly yields the WKI category of NLEEs:
\beq
\left(\begin{array}{c}
q\\
r
\end{array}\right)_t & = & J\cdot G_m, \ \ m\in Z, \label{6.4.13}\\
    G_m & = & \left\{\begin{array}{ll}
{\cal L}^m\cdot(A^{(-2)},B^{(-2)})^T, & m=0,1,2\ldots, \\
{\cal L}^m\cdot(C^{(-2)},D^{(-2)})^T, & m=-1,-2,\ldots,
\end{array}\right. \label{6.4.14}  
\eeq
where $J,\ {\cal L}$ and ${\cal L}^{-1}$ are defined
by (\ref{6.4.3}), (\ref{6.4.4}) and (\ref{6.4.9}), respectively.


   For any given $G=(G^{[1]},G^{[2]})^T \in {\cal B}^2$,
    the equation 
    $[V,L]=L_{*}(K\cdot G)L^{-1}-L_{*}(J\cdot G)$ has the following
    operator solution
\be
  V=V(G)=\left(\begin{array}{lr}
0 & \bar{B}\\
\bar{C} & 0
\end{array}\right)+\bar{A}
\left(\begin{array}{lr}
-i & q\\
r & i
\end{array}\right)L  \label{6.4.15}
\ee
where $\bar{A},\ \bar{B}, \ \bar{C}$ are the following three functions given by
\begin{eqnarray*}
\bar{A}=\bar{A}(G) &=& \frac{1}{2p}
 \left(\frac{q}{p}G_{xx}^{[1]}-\frac{r}{p}G_{xx}^{[2]}\right)^{(-1)}, \ \ p=\sqrt{1-qr}, \\
\bar{B}=\bar{B}(G) &=& \frac{1}{4i}\left(2G_{xx}^{[2]}
   -\pa\frac{q}{p}\cdot \left(\frac{q}{p}G_{xx}^{[1]}-\frac{r}{p}G_{xx}^{[2]}\right)^{(-1)}\right),\\
\bar{C}=\bar{C}(G) &=& \frac{1}{4i}\left(2G_{xx}^{[1]}+\pa\frac{r}{p}\cdot
        \left(\frac{q}{p}G_{xx}^{[1]}-\frac{r}{p}G_{xx}^{[2]}\right)^{(-1)}\right).
\end{eqnarray*}


    Thus, the WKI category (\ref{6.4.13}) has
    the GLR:
\beq
  L_t & = & \left[W_{m}, L\right] + \bar{M}L^{m+1}, \
                   m\in Z,\label{6.4.16}\\
           & & \bar{M}=\left\{\begin{array}{ll}
    \frac{1}{1-qr}\left(\begin{array}{cc}
     qA & iB\\
    iA & -rB
     \end{array}\right), & m\ge0, \\
    \frac{1}{1-qr}\left(\begin{array}{ll}
      qC & iD\\
      iC & -rD
      \end{array}\right), & m<0,
      \end{array}\right. \label{6.4.17}
\eeq
with  the GLO
\be
W_{m}  = \sum V(G_{j})L^{m-j}, \ m\in Z.
   \label{6.4.18}
\ee
Here $L,\ L^{-1}$ and $V(G_j)$ are given by (\ref{6.4.1}), (\ref{6.4.6})
and (\ref{6.4.15}) with $G=G_j$ defined by
(\ref{6.4.14}), respectively.

$ $

{\large \bf 6.5}  \quad  The following spectral problem
\be
L\cdot y=\la y, \ L=L(u)=\frac{1}{u}\left(\begin{array}{cc}
i  & 1-u\\
1 & -i
\end{array}\right)\pa, \ \ y=\left(\begin{array}{l}
y_1\\
y_2
\end{array}\right),  \label{l1}
\ee
yields its Lenard operators pair 
$$K=\partial^{3}, J=-2(\partial u+u\partial).$$
The Gateaux derivative operator $L_{*}(\xi)$ of the spectral operator $L$
in the direction $\xi\in {\cal B}$ is
\begin{equation}
  L_{*}(\xi)=\frac{\xi}{u^{2}}\left(\begin{array}{cc}
-\mit i \rm & -1\\
-1 & \mit i \rm
\end{array}\right)\partial=\frac{\xi}{u}\left(\begin{array}{cc}
0 & -\mit i \rm \\
0 & -1
\end{array}\right)L. \label{l*1}
\end{equation}
Apparently, $L_{*}$ is a homomorphism and
$L_{*}(\xi)=0 \Longleftrightarrow \xi=0$.

In the category derived from Eq. (\ref{l1}), 
we can obtain the Harry-Dym hierarchy
as well as some new
integrable equations. For example, 
the following nonlinear equation
 \beq v_{xt_{-2}}=2vv_{xx}+v_{x}^2 \eeq
is a new integrable
equation with many unknown physical properties.
 In fact, this equation
is included in an isospectral ($\la_{t_{-2}}=0$)
negative order  hierarchy of (\ref{l1}), and
its standard Lax operator is 
$$W_{-2}=-V(G_{-2})L^{-1}
-V(G_{-1})L^{-3},$$
 where $V(G_j)\ (j=-2,-1)$ is given by 
\be
  V=V(G)=G_{xx}\left(\begin{array}{cc}
0 & 1\\
0 & 0
\end{array}\right) + G_{x}\left(\begin{array}{cc}
1 & -2\mit i \rm\\
0 & -1
\end{array}\right)L + 2G\left(\begin{array}{cc}
-\mit i\rm & u-1\\
-1 & \mit i\rm
\end{array}\right)L^{2}. \label{v1}
\ee
with $G=G_{-2}=-v^{(-1)}, \ G_{-1}=\frac{1}{2}$, respectively, and
$L^{-1}$ is the inverse of $L$, given by 
\begin{equation}
  L^{-1}=\left(\begin{array}{cc}
-i\pa^{-1} & \pa^{-1}v_x-\pa^{-1} \\
-\pa^{-1} & i\pa^{-1}
  \end{array}\right). \label{l-1}
\end{equation}

We will give in detail some reductions for the latter four spectral
problems in a later paper.

\section*{\Large \bf Acknowledgments}
The first author (Z. Qiao) would like to express his sincere thanks to
Prof. Gu Chaohao and Prof. Hu Hesheng for their enthusiastic
encouragements and helps, to Prof. Dr. Strampp for his warm invitation,
and also  to D. D. Holm,
J. M. Hyman and L. G. Margolin  for their friendly academic discussions.
The authors are very grateful
to the referees for mentioning  the Refs.  \cite{C1, MSS1,OS1,Dickey1}
and their precious opinions.

This work was supported by the U.S. Department of Energy under
contracts W-7405-ENG-36 and the Applied Mathematical Sciences Program
KC-07-01-01; the Alexander von Humboldt
Foundation, Germany;  the Special Grant of National
Excellent Doctorial Dissertation of China; and also the Doctoral
Programme Foundation of the Insitution of High Education of China.

\end{document}